\documentclass[reprint,aps,prl,showpacs,floatfix,superscriptaddress,citeautoscript,longbibliography]{revtex4-1}

\pdfoutput=1

\usepackage{amsmath, amsfonts, amssymb, amsbsy}
\usepackage{graphicx}
\usepackage{xspace}

\usepackage{pdfpages}

\usepackage[caption=false]{subfig}

\makeatletter
\def\tagform@#1{\maketag@@@{\ignorespaces#1\unskip\@@italiccorr}}
\makeatother

\usepackage{hyperref}
\hypersetup{
        pdftitle={On the nature of the band inversion and the topological phase transition in (Pb,Sn)Se},
        pdfdisplaydoctitle=true,
        pdfsubject = {Article},
        pdfauthor= {BMW},
        pdfkeywords= {},
        colorlinks=true,
        linkcolor=blue,
        citecolor=blue,
        urlcolor=blue,
        unicode=false
}

\newcommand{\PbSnXSe}{Pb\texorpdfstring{$_{1-x}$}{\ifpdfstringunicode{\unichar{"2081}\unichar{"208B}\unichar{"2093}}
{1-x}}Sn\texorpdfstring{$_{x}$}{\ifpdfstringunicode{\unichar{"2093}}{x}}Se\xspace}

\newcommand{\xrange}{\texorpdfstring{$0 \leqslant x \leqslant 0.37$\xspace}
{\ifpdfstringunicode{\unichar{"0030}\unichar{"00A0}\unichar{"2264}\unichar{"00A0}\unichar{"0078}\unichar{"00A0}
\unichar{"2264}\unichar{"00A0}\unichar{"0030}\unichar{"002E}\unichar{"0033}\unichar{"0037}}{0 <= x <= 0.37}}}

\begin{document}
\title{On the nature of the band inversion and the topological phase transition in (Pb,Sn)Se}

\author{B.~M. Wojek}
\homepage{http://bastian.wojek.de/}
\affiliation{KTH Royal Institute of Technology, ICT MNF Materials Physics, Electrum 229, 164 40 Kista, Sweden}
\author{P.~Dziawa}
\affiliation{Institute of Physics, Polish Academy of Sciences, Aleja Lotnik\'{o}w 32/46, 02-668 Warsaw, Poland}
\author{B.~J. Kowalski}
\affiliation{Institute of Physics, Polish Academy of Sciences, Aleja Lotnik\'{o}w 32/46, 02-668 Warsaw, Poland}
\author{A.~Szczerbakow}
\affiliation{Institute of Physics, Polish Academy of Sciences, Aleja Lotnik\'{o}w 32/46, 02-668 Warsaw, Poland}
\author{A.~M. Black-Schaffer}
\affiliation{Department of Physics and Astronomy, Uppsala University, Box 516, 751 20 Uppsala, Sweden}
\author{M.~H.~Berntsen}
\altaffiliation{Present address: Deutsches Elektronen-Synchrotron (DESY), Photon Science, Coherent X-ray Scattering, 
Notkestrasse 85, 22607 Hamburg, Germany}
\affiliation{KTH Royal Institute of Technology, ICT MNF Materials Physics, Electrum 229, 164 40 Kista, Sweden}
\author{T.~Balasubramanian}
\affiliation{MAX IV Laboratory, Lund University, P.O. Box 118, 221 00 Lund, Sweden}
\author{T.~Story}
\affiliation{Institute of Physics, Polish Academy of Sciences, Aleja Lotnik\'{o}w 32/46, 02-668 Warsaw, Poland}
\author{O.~Tjernberg}
\email{oscar@kth.se}
\affiliation{KTH Royal Institute of Technology, ICT MNF Materials Physics, Electrum 229, 164 40 Kista, Sweden}

\date{\today}

\begin{abstract}
The recent discovery of a topological phase transition in IV-VI narrow-gap semiconductors has revitalized the 
decades-old interest in the bulk band inversion occurring in these materials. Here we systematically study the (001) 
surface states of \PbSnXSe mixed crystals by means of angle-resolved photoelectron spectroscopy in the parameter space 
\xrange and $300~\text{K}\geqslant{}T\geqslant{}9~\text{K}$. Using the surface-state observations, we monitor directly 
the topological phase transition in this solid solution and gain valuable information on the evolution of the 
underlying fundamental band gap of the system. 
In contrast to common model expectations, the band-gap evolution appears to be nonlinear as a function of the studied 
parameters, resulting in the measuring of a discontinuous band inversion process. 
This finding signifies that the anticipated gapless bulk state is in fact not a stable configuration and that the 
topological phase transition therefore exhibits features akin to a first-order transition.

\end{abstract}


\maketitle

\section{Introduction}
\label{sec:intro}

Lead chalcogenides and related compounds have been studied intensely already throughout the last 
century~\cite{NimtzAndSchlicht, SpringholzBauer-2014}. These narrow-gap semiconductors are used for applications in 
infrared lasers~\cite{Preier-ApplPhys-1979} and detectors~\cite{Rogalski} as well as in thermoelectric 
devices~\cite{Harman-JElectronMater-2000, Heremans-EnergyEnvironSci-2012}. They also exhibit a large range of peculiar 
fundamental properties, such as positive temperature and negative pressure coefficients of the energy 
gap~\cite{Schlueter-PhysRevB-1975} and nonparabolic band dispersions~\cite{Mitchell-PhysRev-1966}. Moreover, for 
suitable compounds and parameter ranges, the band gap undergoes an inversion as a function of temperature, 
pressure~\cite{Martinez-PhysRevB-1973-I, *Martinez-PhysRevB-1973-II, *Martinez-PhysRevB-1973-III}, and 
composition~\cite{Dimmock-PhysRevLett-1966, Strauss-PhysRev-1967}. At the interface between inverted and noninverted 
insulating layers gapless states were predicted to form~\cite{Volkov-JETPLett-1985}. The interest in this class of 
materials was very recently renewed by the investigation of the so-called topological-crystalline-insulator (TCI) state 
in SnTe~\cite{Hsieh-NatCommun-2012, Tanaka-NatPhys-2012} as well as the solid solutions 
(Pb,Sn)Te~\cite{Xu-NatCommun-2012} and (Pb,Sn)Se~\cite{Dziawa-NatMater-2012}. In this state of matter, the mirror 
symmetry present in the rock-salt structure ensures degenerate energy eigenvalues along mirror lines, and hence metallic 
surface states on certain high-symmetry surfaces~\cite{Safaei-PhysRevB-2013, *Liu-PhysRevB-2013}, when the band gap is 
inverted. Additionally, these topologically protected surface states exhibit Dirac-like dispersions and they are 
spin-momentum-locked~\cite{Xu-NatCommun-2012, Wojek-PhysRevB-2013, Wang-PhysRevB-2013}.

Although the tunable band inversion in the lead and tin chalcogenides has been established for decades, the details of 
the gap closing remain in the dark. It is commonly expected (cf. e.g. most of the aforementioned references) that the 
size of the band gap goes to zero when a critical composition, temperature or pressure is reached. Yet, 
a \emph{completely closed} band gap is naturally hard to observe experimentally. The lowest confirmed gap 
values are in the region of a few tens of millielectronvolts. For instance, in infrared absorption studies, the 
long-wavelength limit was not accessible~\cite{Strauss-PhysRev-1967} and very-low-energy laser emission seems to be 
hindered by plasmon-phonon excitations~\cite{Harman-ApplPhysLett-1969, *Calawa-PhysRevLett-1969, 
Martinez-PhysRevB-1973-I, *Martinez-PhysRevB-1973-II, *Martinez-PhysRevB-1973-III}.

The recent discovery of the TCI phase and the appertaining surface states has opened a new route to study the details 
of the band inversion in this class of materials by high-resolution photoelectron spectroscopy. The (001) 
surface states at the surface high-symmetry point $\overline{X}$ lie within the bulk band gap just beyond the valence- 
and conduction-band edges projected from the bulk $L$ points~\cite{Hsieh-NatCommun-2012, Dziawa-NatMater-2012, 
Wojek-PhysRevB-2013, Barone-PhysRevB-2013}. The same is true for the corresponding high-symmetry points of the (110) 
surface~\cite{Safaei-PhysRevB-2013, *Liu-PhysRevB-2013}. Hence, studying the surface-state evolution across a parameter 
range for which the band inversion occurs, can give a reliable estimate of the band gap at $L$ and thus further 
elucidate the process of the inversion.

In this article, we report on a systematic angle-resolved photoelectron spectroscopy (ARPES) study of 
(001)-oriented (Pb,Sn)Se mixed crystals. The topological phase boundary is established by the investigation of 
the evolution of the surface states around $\overline{X}$ as a function of composition and temperature. 
Moreover, contrary to the common theoretical expectations and the conclusions from previous experimental work, our 
observations point to an unstable zero-gap state (ZGS) in the bulk material.

\section{Experimental details and results}

\begin{figure*}
\centering
\includegraphics[width=\textwidth]{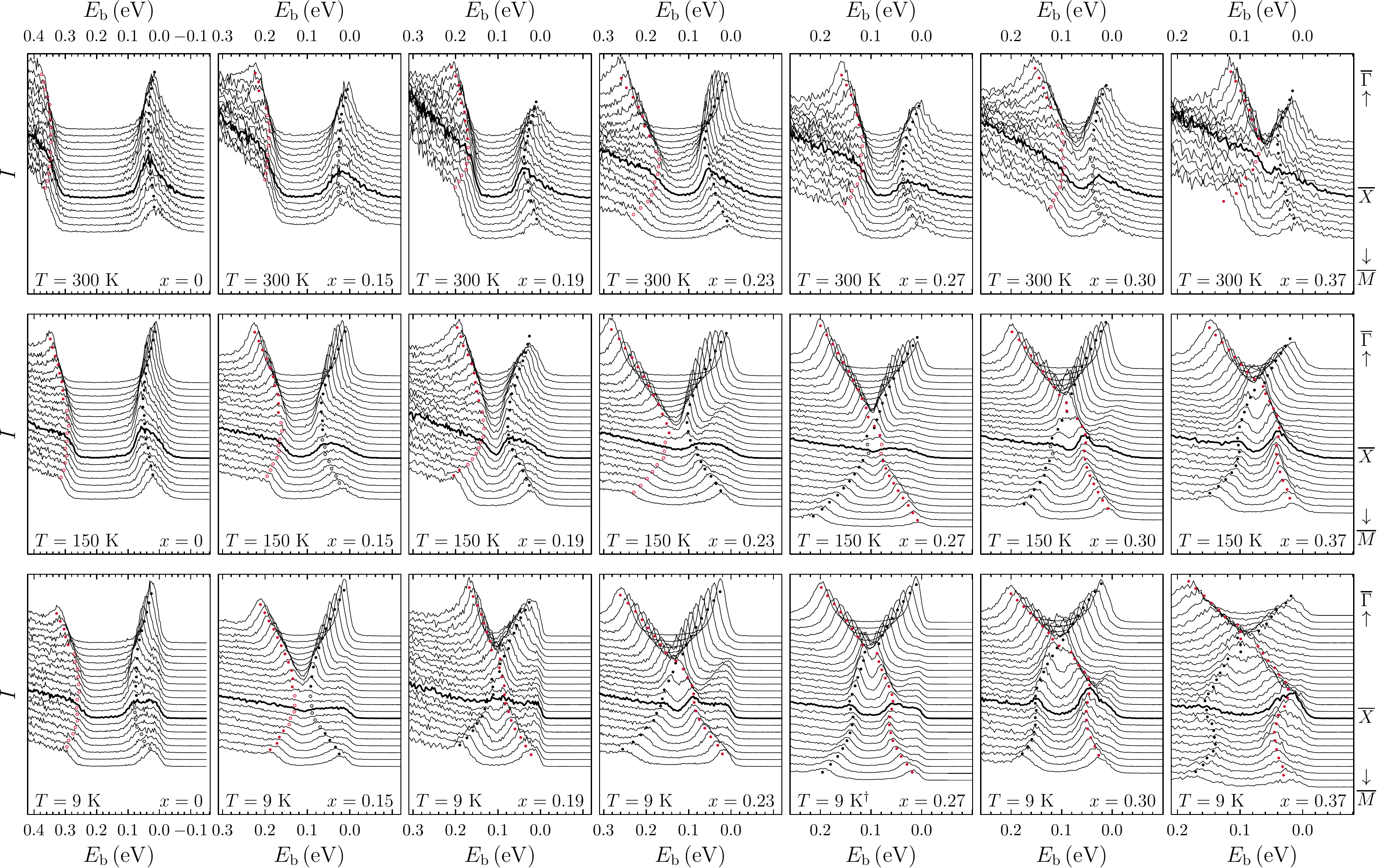}
\caption{ARPES spectra in the vicinity of $\overline{X}$ of the (001) surface of \PbSnXSe as a function of tin 
content $x$ (varying with the columns) for three selected temperatures~\cite{NatMat-correction, Supplement}. The 
observed surface states display a transition from an overall gapped state (low $x$ or high $T$)~\cite{BandCurvature} to 
a state that is gapless along the $\overline{\Gamma}$-$\overline{X}$ line (high $x$ and low $T$)~\cite{LowTempGap}. The 
energy-distribution-curve (EDC) intensity is shown on a linear scale and the spacing between neighboring EDCs is roughly 
given by $0.009\,\pi/a$, where $a(x,T)$ is the bulk lattice constant. In each panel, the 
solid circles mark fitted Voigtian EDC peak positions; where no peaks are discernible open circles mark the 
band edge. The latter positions are generally found to extrapolate the dispersion of the surface-state peak positions 
smoothly. The different colors are merely used to visualize the derived $L_6^-$ and $L_6^+$ character of the valence- 
and conduction-band edges.}
\label{fig:matrix}
\end{figure*}

The \PbSnXSe (\xrange) single crystals used in this study have been grown by the self-selecting vapor-growth 
method~\cite{Szczerbakow-JCrystalGrowth-1994, *Szczerbakow-ProgCrystalGrowth-2005}. Their composition has been 
determined by means of X-ray diffraction as well as energy-dispersive X-ray spectroscopy. Powder X-ray diffraction 
confirmed the rock-salt structure [space group $Fm\overline{3}m$ (225)] both at room temperature and $T=15$~K. To avoid 
the necessity of extrinsic surface doping~\cite{Pletikosic-PhysRevLett-2014} which might alter the intrinsic 
electronic structure, bulk $n$-type crystals are used in this study. The ARPES measurements in the temperature range 
$300~\text{K}\geqslant{}T\geqslant{}9~\text{K}$ have been conducted on samples cleaved along a (001) surface in 
ultra-high vacuum. The experiments have been performed using the laser-based ARPES set-up \textsc{baltazar} equipped 
with a time-of-flight electron analyzer~\cite{Berntsen-RevSciInstrum-2011}. Linearly polarized light with a photon 
energy $h\nu=10.5$~eV was used to excite the electrons. The total energy and crystal-momentum resolution was about 
$5$~meV and better than $0.008$~\AA{}$^{-1}$, respectively.

Characteristic ARPES spectra are shown in Fig.~\ref{fig:matrix}. The data in the vicinity of $\overline{X}$ are plotted 
along the high-symmetry lines of the surface Brillouin zone ($\overline{\Gamma}$-$\overline{X}$-$\overline{M}$). In all 
spectra surface states are discernible. While the samples at high temperature ($T = 300$~K) exhibit gapped states for 
all $x$~\cite{BandCurvature}, at lower temperatures the spectra show a change from gapped surface states at 
low $x$ to metallic (gapless) states crossing on the $\overline{\Gamma{}}$-$\overline{X}$ line at high $x$---the 
hallmark of the transition from a usual band insulator to a TCI~\cite{LowTempGap}. A closer inspection of the spectra 
shows that for $x\leqslant 0.15$, the energy gap at $\overline{X}$ ($\Delta_{\overline{X}}$) decreases when the 
temperature is lowered. The gap $\Delta_{\overline{X}}$ also shrinks with increasing $x$ when the samples are kept at 
room temperature. The 
opposite behavior is found for samples with high $x$ at low temperatures.

As described in the introduction above, the surface-state dispersion can be viewed as an envelope of the projected bulk 
bands. Hence, $\Delta_{\overline{X}}$ provides an estimate of the bulk band gap at 
$L$~\cite{Hsieh-NatCommun-2012, Dziawa-NatMater-2012, Wojek-PhysRevB-2013, Barone-PhysRevB-2013} and thus, the observed 
qualitative trends are generally anticipated. They can be explained with the negative pressure coefficient of the energy 
gap (the lattice constant decreases with temperature as well as with increasing $x$) together with the phonon-related 
changes (cf. Ref.~\onlinecite{NimtzAndSchlicht} and references therein). Also in line with the expectations is that for 
PbSe the top of the valence band changes more strongly than the bottom of the conduction band across the range of 
parameters---a direct consequence of the band repulsion due to the occupied Pb $6s$ level in the 
valence band~\cite{Schlueter-PhysRevB-1975, Wei-PhysRevB-1997}.

\section{Analysis and discussion}

\begin{figure}
\centering
\begin{minipage}[c]{0.485\columnwidth}
  \vspace{0pt}\subfloat{\includegraphics[width=\textwidth]{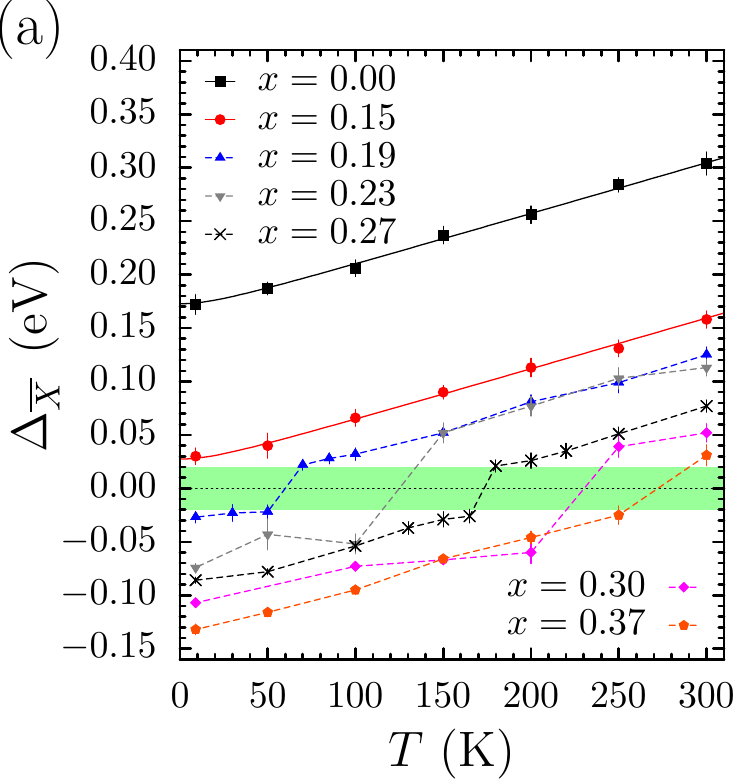}\label{fig:gap:a}}\\[2mm]
              \subfloat{\includegraphics[width=\textwidth]{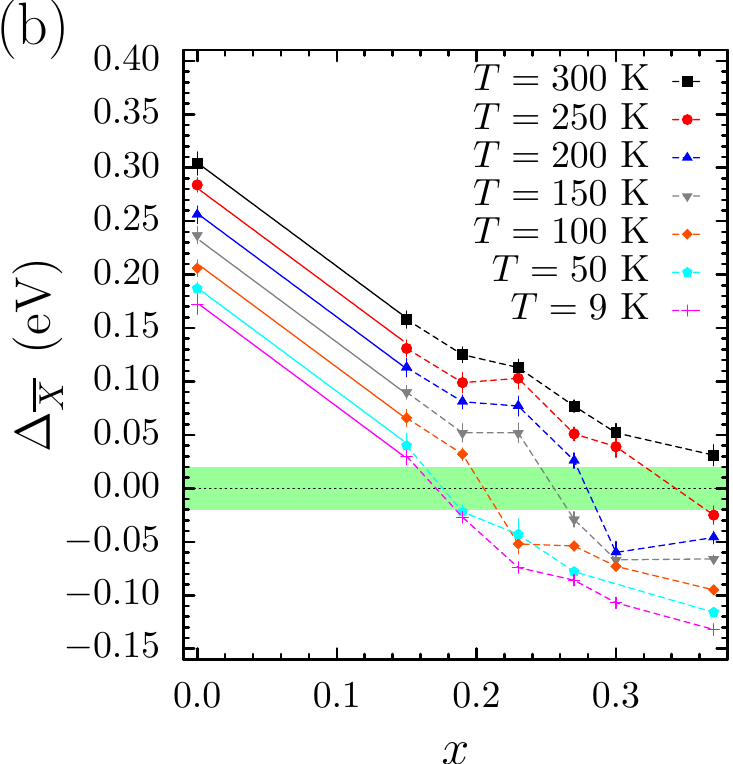}\label{fig:gap:b}}
\end{minipage}
\hfill
\begin{minipage}[c]{0.495\columnwidth}
  \vspace{0pt}\subfloat{\includegraphics[width=\textwidth]{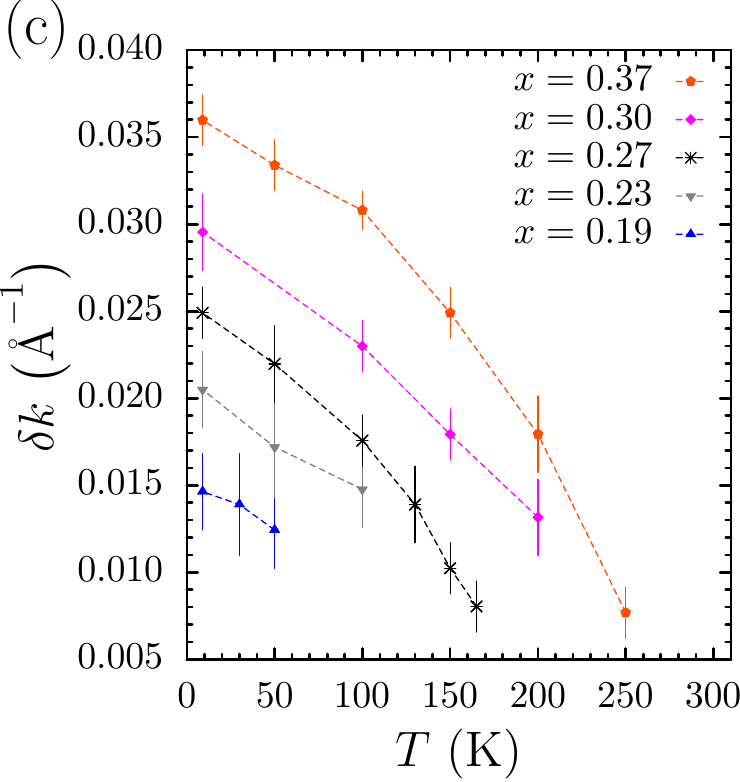}\label{fig:gap:c}}\\[2mm]
              \subfloat{\includegraphics[width=\textwidth]{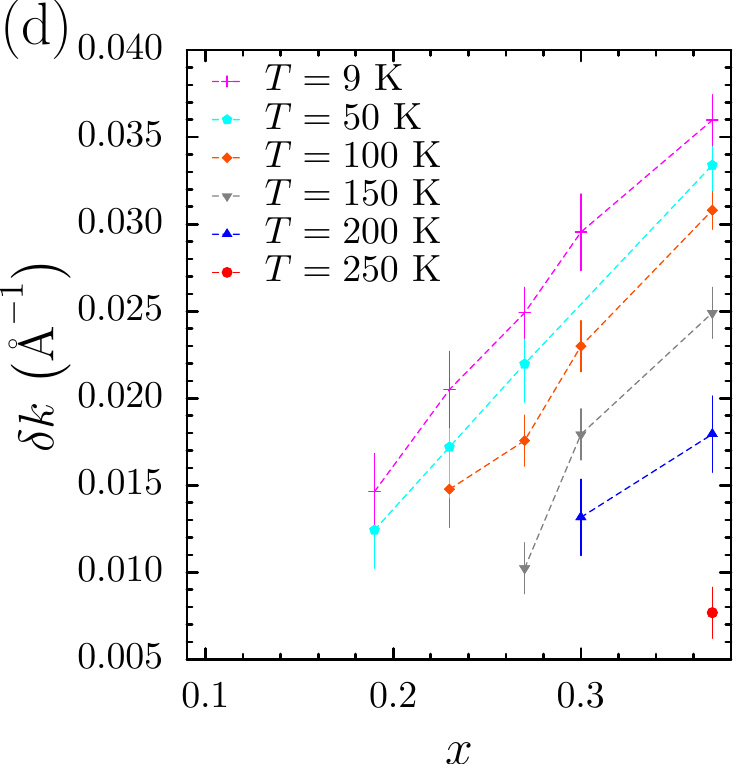}\label{fig:gap:d}}
\end{minipage}
\caption{(a) Energy gap at $\overline{X}$ as a function of temperature for \PbSnXSe crystals with different tin contents 
$x$. The gap values have been determined from ARPES spectra as the ones shown in 
Fig.~\ref{fig:matrix}~\cite{Supplement}. In the TCI state, $\Delta_{\overline{X}}$ is taken to be negative. The 
green-shaded area represents the minimal gap that is always observed. The solid lines represent fits of 
Eq.~\eqref{eq:Grisar} to the data. The dashed lines are merely guides to the eye. (b) Same data as in (a) but presented 
as a function of $x$. (c) Distance between $\overline{X}$ and 
the surface-state band crossing in the TCI phase. (d) Same data as in (c) but presented as a 
function of $x$. The dashed lines are guides to the eye.}
\label{fig:gap}
\end{figure}

To gain a more quantitative insight in the evolution of the energy gap, in particular in the region where the gap is 
small, we plot the values of $\Delta_{\overline{X}}(x,T)$ in Fig.~\ref{fig:gap}. The shown values represent the 
difference in the surface-state positions at the band edges determined by the local extrema in the dispersion along the 
high-symmetry directions. The determined band-edge positions are marked in Fig.~\ref{fig:matrix}~\cite{Supplement}. We 
correlate the formation of a metallic surface state with an inverted band structure which, following the usual 
convention, has a negative energy gap. The resulting temperature and tin-content dependencies are shown in 
Figs.~\ref{fig:gap:a} and~\ref{fig:gap:b}, respectively. Concentrating first on the low-$x$ samples ($x \leqslant 
0.15$), we see that the band gap evolves nearly linearly as expected and commonly accepted. The data are very well 
described by the phenomenological model~\cite{Grisar}
\begin{equation}
\Delta = E_0 + \alpha\cdot{}x + \sqrt{E_1^2 + \left(\beta\cdot{}T\right)^2}.
\label{eq:Grisar}
\end{equation}
The values of the parameters are given in Tab.~\ref{tab:parameters} and they are overall compatible with those found in 
literature. However, when turning to the high-$x$ samples, it is apparent immediately that Eq.~\eqref{eq:Grisar} does 
not hold anymore when the band gap becomes small and inverts. The curves for $x \geqslant 0.19$ all show a 
discontinuous 
band inversion, irrespective of the critical temperature [Fig.~\ref{fig:gap:a}] or the critical composition 
[Fig.~\ref{fig:gap:b}]. Moreover, the minimal observed absolute value of $\Delta_{\overline{X}}$ is about ($20$ 
to $25$)~meV and therefore slightly less but of the same order of magnitude as the smallest obtained laser 
energy~\cite{Harman-ApplPhysLett-1969}. Even though deviations from the approximately linear gap evolution were pointed 
out early on~\cite{Harman-ApplPhysLett-1969, Grisar}, models of type~\eqref{eq:Grisar} are still widely in use and also 
stable gapless bulk states are yet postulated~\cite{Svane-PhysRevB-2010}.

\begin{table}
\caption{\label{tab:parameters} Parameters used in Eq.~\eqref{eq:Grisar}.}
\begin{ruledtabular}
\begin{tabular}{ccccc}
$E_0$~(meV) & $E_1$~(meV) & $\alpha$~(meV) & $\beta$~(meV/K) & Ref.\tabularnewline
\noalign{\smallskip}
\hline
\noalign{\smallskip}
$130$ & $0$ & $-890$ & $0.45$ & \cite{Strauss-PhysRev-1967}\tabularnewline
$125$ & $20$ & $-1021$ & $0.506$ & \cite{Grisar}\tabularnewline
$161(7)$ & $12(9)$ & $-969(20)$ & $0.477(27)$ & this work\tabularnewline
\end{tabular}
\end{ruledtabular}
\end{table}

Before discussing the observation of the discontinuous inversion further, we would like to point out that the 
determination of the position of the gapped surface states in the normal-insulator case is somewhat hindered by the 
very low spectral weight \emph{at} $\overline{X}$, where these states merge into the projected bulk states (cf. 
Fig.~\ref{fig:matrix} and Ref.~\onlinecite{Barone-PhysRevB-2013}). While this creates some uncertainty in the 
$\Delta_{\overline{X}}$ values in the normal state, our estimates---based on the continuous bands having a finite 
curvature---are overall conservative. Additionally, it appears as if occasionally the bottom of the 
conduction-band surface state is shifted away from $\overline{X}$ reminiscent of a ``Rashba-like-split'' surface state. 
Although, we cannot exclude entirely, that the shift in the normal band-insulator phase is merely an artifact of the 
missing spectral weight at $\overline{X}$, the rather clearly visible dispersion in the $x=0.23$ high-temperature data 
indicates a real effect that so far cannot be reconciled with model calculations~\cite{Wojek-PhysRevB-2013, 
Barone-PhysRevB-2013, Dziawa-NatMater-2012}. It is worth noting that for (Pb,Sn)Te recent calculations indicate a 
substantial influence of a finite surface potential gradient on the (111) surface states both in the TCI and the normal 
state~\cite{Yan-arXiv-2014}.
Until similar calculations become available for the (001) surface we only further quantify the separation of the Dirac 
points on the mirror line $\overline{\Gamma}$-$\overline{X}$ in the TCI state of (Pb,Sn)Se. As shown in 
Figs.~\ref{fig:gap:c} and~\ref{fig:gap:d} the Dirac points move away farther from $\overline{X}$ in the direction of 
$\overline{\Gamma}$ when the samples ``advance deeper'' into the TCI state, consistent with observations made previously 
for (Pb,Sn)Te~\cite{Tanaka-PhysRevB-2013}.

Despite the above considerations, the observation of a nonlinear band evolution and the discontinuous inversion across 
the whole studied parameter range remains intact. Specifically, a linear gap evolution would suggest the high-$x$ mixed 
crystals to be found in the TCI phase at room temperature. Yet, overall gapped surface states are observed. In 
addition, 
also published ARPES data on (Pb,Sn)Te suggest a similar discontinuous band inversion [cf. Fig.~3(d) of 
Ref.~\onlinecite{Tanaka-PhysRevB-2013}]. So what is responsible for observing an open bulk band gap throughout all 
measurements? It seems that pure lattice dilatation effects play no role here, since computational studies of the 
pressure and strain dependence of the band gap and the topological transition in binary compounds yield a continuous 
basically linear inversion with a zero crossing of the gap~\cite{Barone-PhysRevB-2013, Niu-MaterialsExpress-2013}. 
However, when changing the temperature, the lattice dilatation constitutes only about half of the gap variation in the 
lead salts~\cite{Schlueter-PhysRevB-1975}. The other half is attributed to electron-phonon 
interactions~\cite{Keffer-PhysRevB-1970}. Hence, lattice vibrations (and also mixed plasmon-phonon modes) are generally 
a grave factor when determining the accurate electronic structure in the studied materials and inter-band scattering 
becomes potentially important for small gap sizes.
For degenerate semiconductors, as they are studied here, also carrier-carrier scattering plays a role in the transport 
properties~\cite{Ravich-JPhysColloques-1968}. Therefore, interaction effects might not be entirely negligible. Finally, 
we emphasize that the nature of the band inversion is studied here in a solid solution. In such compounds disorder is 
always present and influences the electronic properties, although it does not hinder the transition into the TCI 
phase~\cite{Fu-PhysRevLett-2012}. For example, ``alloy scattering'' is known to reduce the overall carrier mobility in 
mixed crystals compared to pure binary materials~\cite{Martinez-PhysRevB-1973-II}. 
Only recently, an effort has been made to study the band inversion in (Pb,Sn)Te solid solutions by \emph{ab initio} 
methods~\cite{Gao-PhysRevB-2008}. It was found that fully ordered structures exhibit semimetallic behavior, but that 
introducing short-range disorder leads to the formation of semiconductor band gaps, although still smaller than 
experimentally measured. Going beyond short-range disorder is, however, very computationally challenging.

Here, we propose a scenario for a band-inversion process in which the ZGS is essentially an unstable 
configuration for the system. The ZGS can generally be described by a band degeneracy forming a Dirac point.
It is a general property of a Dirac point that, in the absence of symmetries protecting it, the spectrum 
easily becomes gapped~\cite{Wehling-AdvPhys-2014}. In the bulk of \PbSnXSe there are no symmetries disallowing having 
a finite bulk gap at any value of the external parameters, such as $x$ or $T$, even around the topological phase 
transition (TPT). As a consequence, although the bulk band gap necessarily has to be zero at some point, 
when evolving from a positive (topologically trivial state) to a negative (topologically non-trivial state) band gap, 
the ZGS itself can be essentially unstable and thus effectively not detectable in any measurement.

To demonstrate the instability of the ZGS within a simple model we use $\boldsymbol{k} \cdot \boldsymbol{p}$ theory 
near 
the bulk band gap minima at any of the $L$ points. The low-energy bulk Hamiltonian can there generally be written 
as~\cite{Mitchell-PhysRev-1966, Hsieh-NatCommun-2012}:
\begin{equation}
\label{eq:H3D}
\mathcal{H} = m\sigma_z + v(k_1 s_2 - k_2 s_1)\sigma_x + v'k_3\sigma_y.
\end{equation}
Here $\boldsymbol{\sigma}$ and $\boldsymbol{s}$ are Pauli matrices, with the eigenvalues of $\sigma_z$ ($\pm 1$) 
labeling the cation (Pb or Sn) or the anion (Se) $p$ orbitals, whereas the spectrum of $s_3$ ($\pm 1$) encodes for the 
Kramers (total angular momentum) degeneracy. The orthogonal coordinate system for the $\boldsymbol{k}$ momentum vector 
has $k_3$ along $\Gamma$-$L$ and $k_1$ aligned with the $[110]$ direction perpendicular to the mirror plane. Moreover, 
the sign of the mass term $m$ determines the topological nature of the material. The normal phase has $m>0$, whereas 
for 
$m<0$ the band structure is inverted and the material is in the nontrivial TCI phase. Note that there cannot be any 
symmetry protecting a ZGS ($m = 0$), since the $m \sigma_z$ mass term is always present in the Hamiltonian, on both 
sides of the TPT.
The energy bands of the low-energy Hamitonian Eq.~\eqref{eq:H3D} are doubly degenerate and given by $E = \pm 
\sqrt{v^2(k_1^2+k_2^2) + v'^2k_3^2 + m^2}$, which forms an anisotropic Dirac spectrum with an energy gap equal to the 
mass $m$. Experimental evidence for massive bulk Dirac fermions in the TCI phase has been reported by recent transport 
measurements~\cite{Liang-NatCommun-2013}.
Using the band structure we calculate the electronic free energy as a function of the mass $m$ for the Hamiltonian in 
Eq.~\eqref{eq:H3D}. We find that the free energy is reduced for a finite $m$ compared to the $m = 0$ spectrum. Thus, 
with no symmetries disallowing a finite $m$ and based on the electronic free energy, a finite energy gap is both allowed 
and energetically favorable. Equation~\eqref{eq:H3D} assumes an intrinsic system with the chemical potential in the 
middle of the bulk band gap. However, we find that adding a finite $n$-type doping, as evident in the experimental 
system, does not qualitatively change the free energy preference of the gapped system.

Thus, unlike in ``true'' zero-gap semiconductors~\cite{Averous-PhysStatSolB-1979}, a perturbation of the system, like 
potentially the aforementioned disorder, phonon and interaction contributions, can lead to the stabilization of a 
fully gapped bulk state. This scenario implies a very sharp, or possibly even first-order, TPT, in which a finite 
energy gap is present in the bulk everywhere except in a very narrow region, where the gap quickly changes sign. In 
fact, the characteristic ``flattening out'' of a first-order transition of the energy gap as function of the tuning 
parameter near the TPT is visible in the ARPES data in Figs.~\ref{fig:gap:a} and~\ref{fig:gap:b}.
First-order transitions between two topologically distinct phases have previously been predicted between a 
topological insulator and a Mott insulating phase in the presence of strong electron-electron 
interactions~\cite{Varney-PhysRevB-2010, Yoshida-PhysRevB-2012}. Our results suggest that sharp, and possibly even 
first-order, TPTs might not just be limited to strongly correlated electron systems.
Thus, even if simple theoretical models, such as tight-binding approaches using the virtual crystal approximation (VCA) 
for solid solutions, seem to capture the coarse features of the band inversion and the 
TPT well~\cite{Dziawa-NatMater-2012, Wojek-PhysRevB-2013, Barone-PhysRevB-2013}, a more accurate description of the 
electronic state of the IV-VI narrow-gap semiconductors in the region where the fundamental band gap is of the order of 
a few tens of millielectronvolts is clearly needed. Such models would need to go beyond a static treatment of the 
crystal lattice and the VCA for solid solutions.

\begin{acknowledgments}
We thank M.~Sahlberg (Uppsala University) and M.~Hudl (KTH Royal Institute of Technology) for supporting us with 
low-temperature powder X-ray diffraction measurements and A.~V. Balatsky for stimulating discussions. This work was 
made possible through support from the Knut and Alice Wallenberg Foundation, the Swedish Research Council, the European 
Commission Network SemiSpinNet (PITN-GA-2008-215368), the European Regional Development Fund through the Innovative 
Economy grant (POIG.01.01.02-00-108/09), and the Polish National Science Centre (NCN) Grant No. 2011/03/B/ST3/02659. 
P.~D. and B.~J.~K. acknowledge the support from the Baltic Science Link project coordinated by the Swedish Research 
Council, VR.
\end{acknowledgments}

\newpage

\onecolumngrid
\linespread{1.5}

\section{Supplemental material}

\setcounter{figure}{0}
\renewcommand{\thefigure}{S\arabic{figure}}
\renewcommand{\theHfigure}{S\arabic{figure}}

\begin{figure*}
\centering
\includegraphics[width=.89\textwidth]{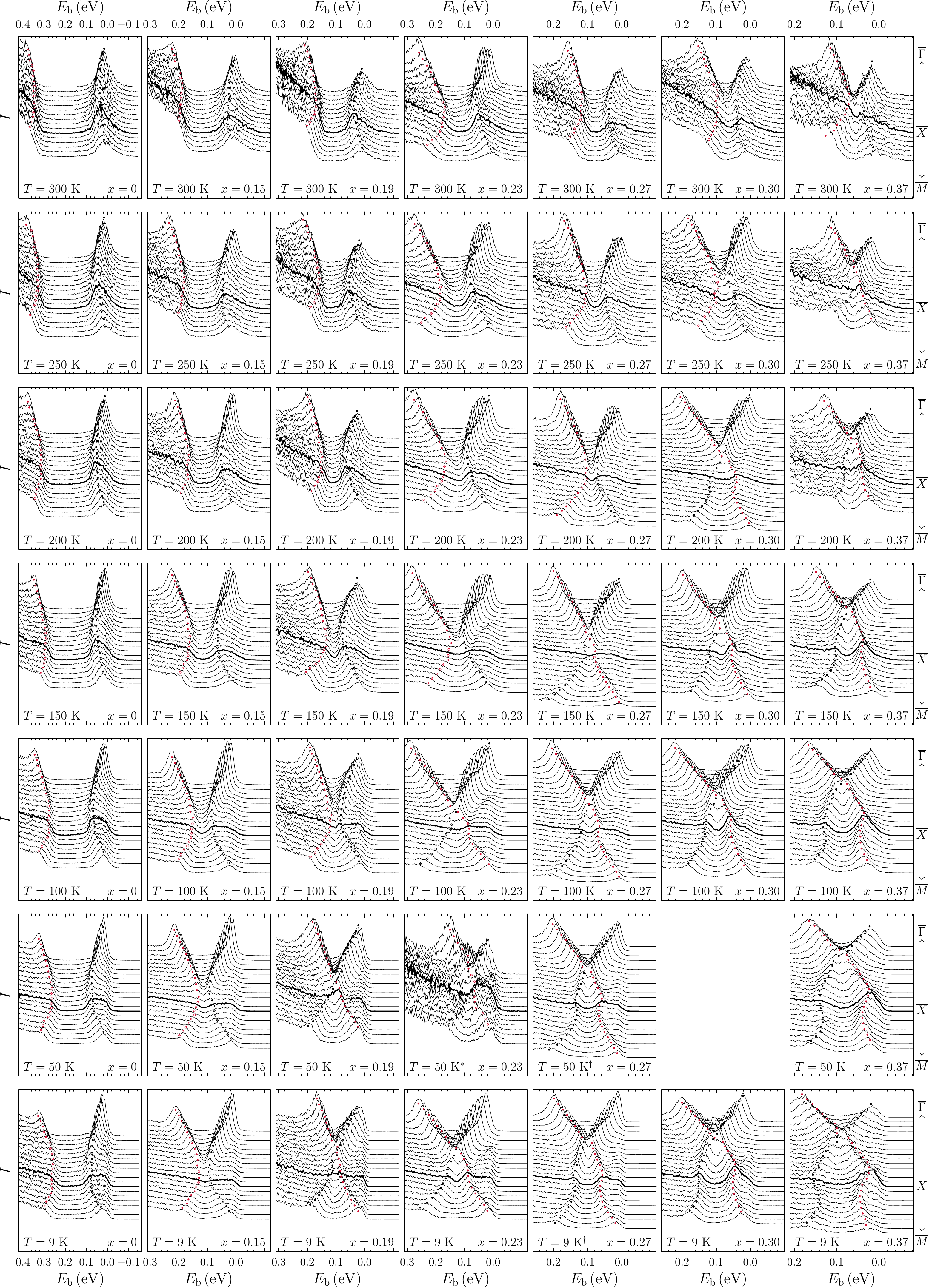}
\caption{ARPES spectra (EDCs) along the high-symmetry lines in the vicinity of $\overline{X}$ of the (001) surface of 
\PbSnXSe as a function of tin content $x$ (varying with the columns) for seven selected temperatures.}
\label{fig:EDCmatrix}
\end{figure*}

\begin{figure*}
\centering
\includegraphics[width=.99\textwidth]{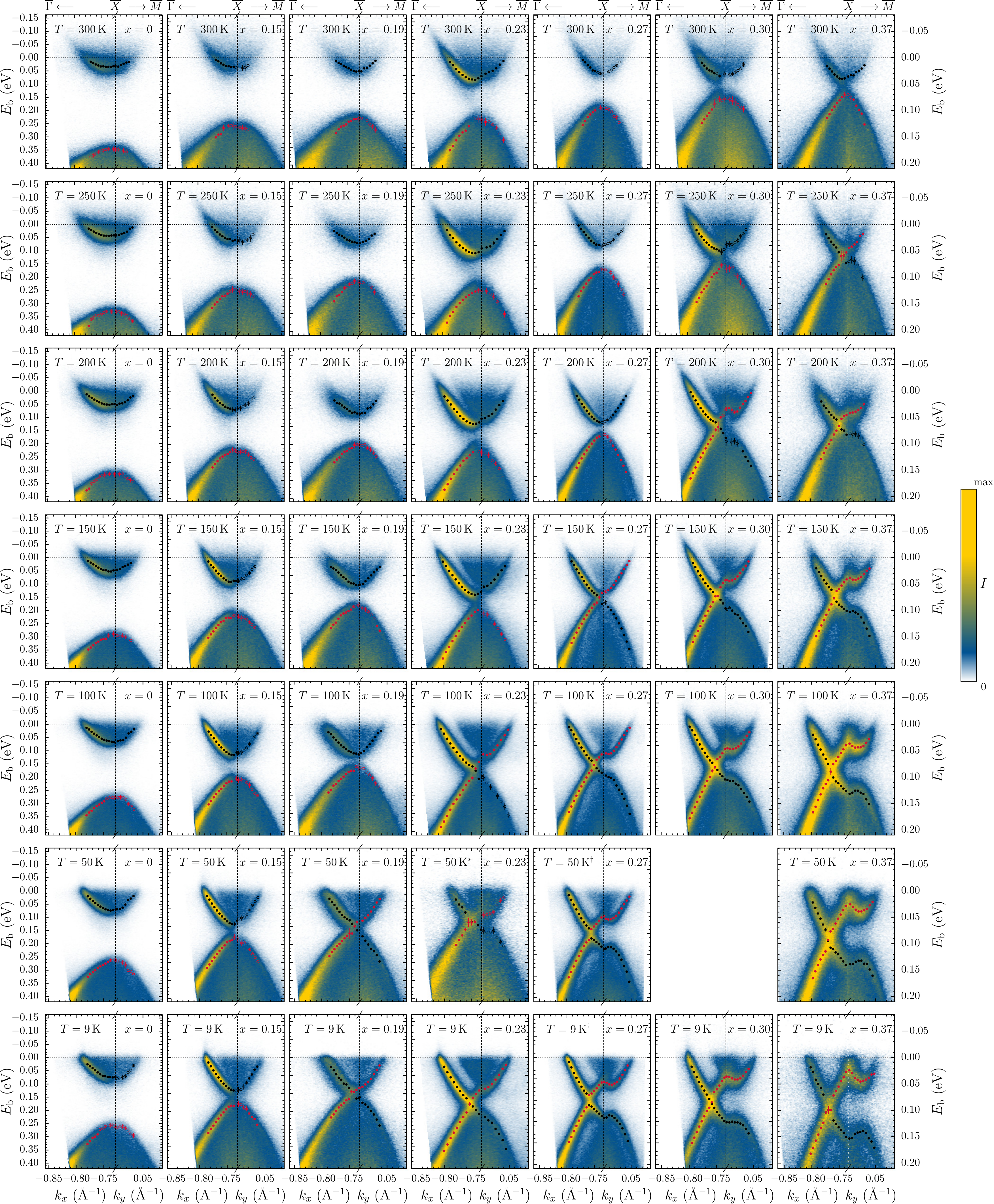}
\caption{Same data as in Fig.~\ref{fig:EDCmatrix} but represented as color plots. Please observe that the energy scale 
changes with $x$. The distance between major (minor) tics on the axes of ordinates is chosen to be always $50$~meV 
($10$~meV) and the Fermi energies ($E_{\mathrm{b}}=0$) are aligned. On the abscissa the $\overline{X}$ positions are 
aligned. Their numerical values vary slightly with the changing lattice parameters.}
\label{fig:COLORmatrix}
\end{figure*}

\begin{figure*}
\centering
\includegraphics[width=.45\textwidth]{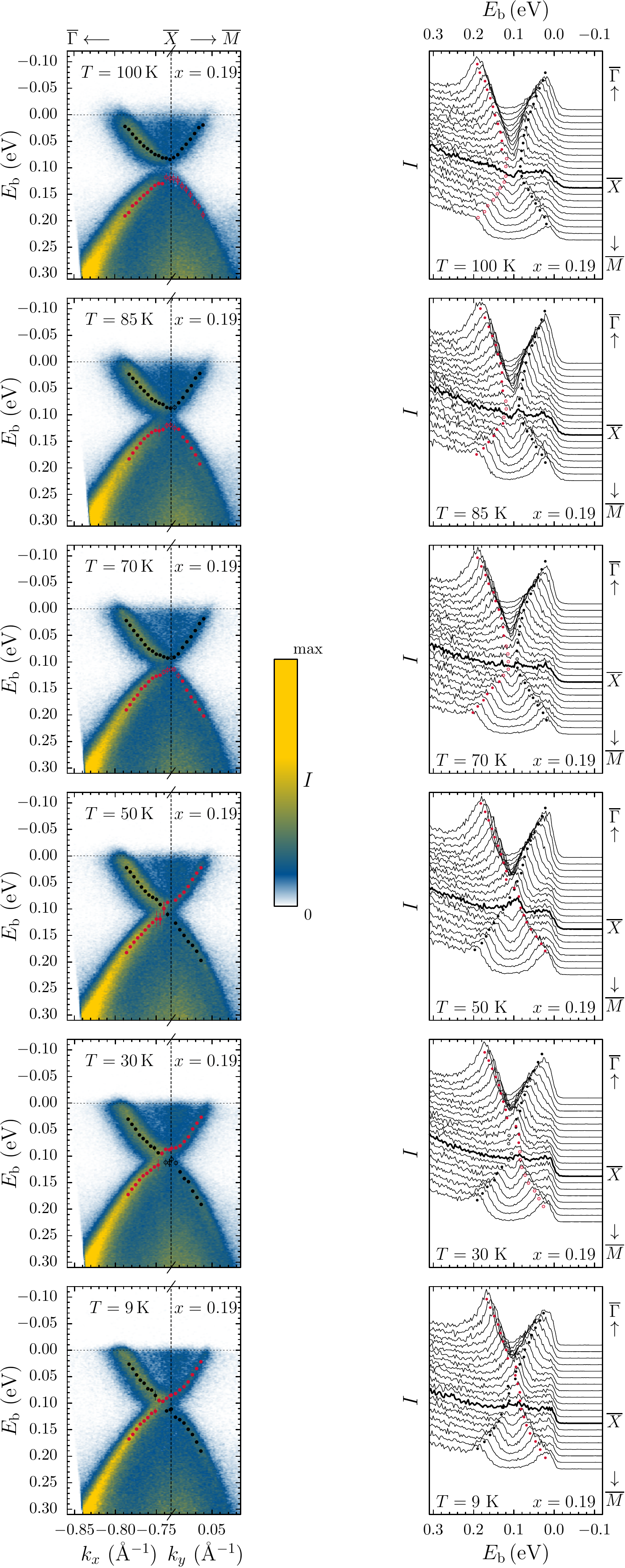}\hfill
\includegraphics[width=.45\textwidth]{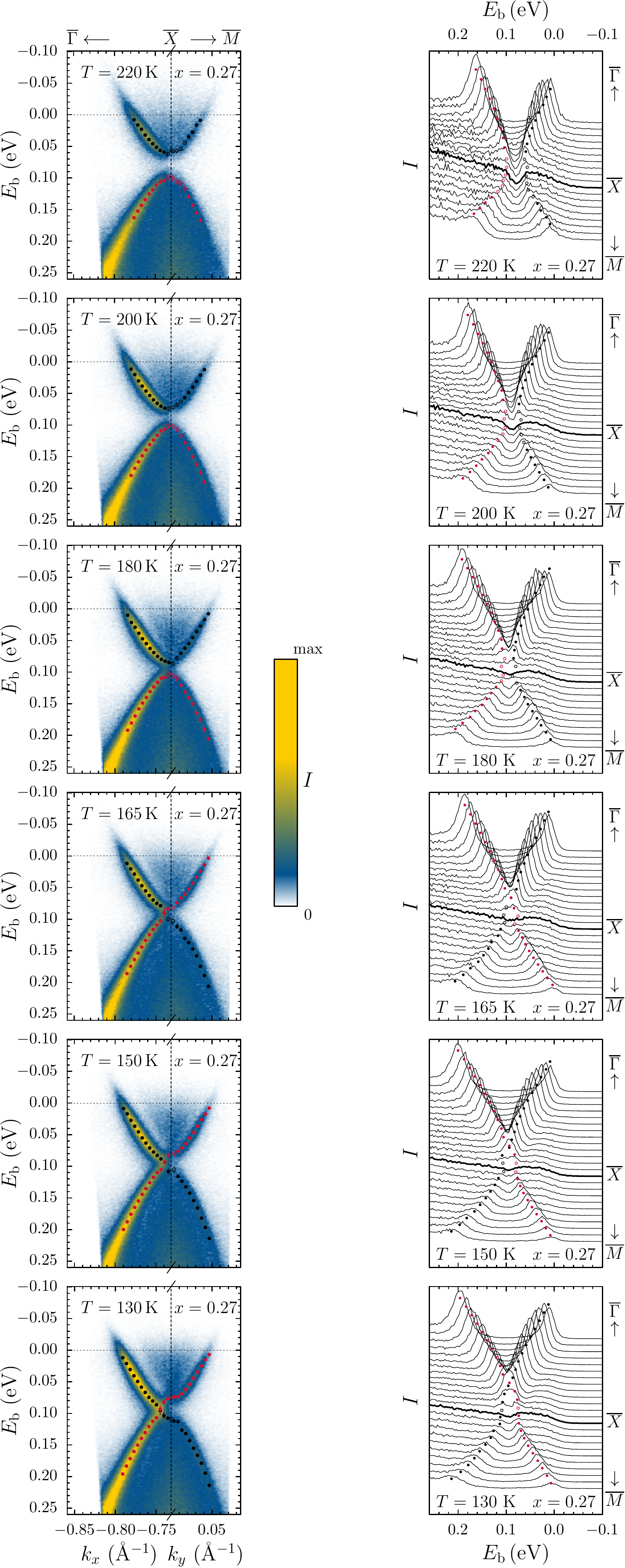}
\caption{ARPES spectra (color plots and corresponding EDCs) along the high-symmetry lines in the vicinity of 
$\overline{X}$ of the (001) surface of Pb$_{0.81}$Sn$_{0.19}$Se (left half of the figure) and Pb$_{0.73}$Sn$_{0.27}$Se 
(right half of the figure) for several temperatures close to the transition into the TCI state.}
\label{fig:X19X27}
\end{figure*}

Figures~\ref{fig:EDCmatrix}, \ref{fig:COLORmatrix}, and \ref{fig:X19X27} included in this document show ARPES spectra 
along the high-symmetry lines in the vicinity of $\overline{X}$ of the (001) surface of \PbSnXSe. While 
Fig.~\ref{fig:matrix} in the main text shows a representative subset of data, here \emph{all} analyzed spectra 
contributing to Fig.~\ref{fig:gap} in the main text are depicted, both as series of energy distribution curves (EDCs) 
and as color plots. The EDC intensity is shown on a linear scale and the spacing between neighboring EDCs is roughly 
given by $0.009\,\pi/a$, where $a$ is the bulk 
lattice constant. The intensity scale of the corresponding color plots is linear as well. In each figure, the solid 
(open) circles mark the determined EDC peak (band edge) positions. The different colors are merely used to visualize 
the derived $L_6^-$ and $L_6^+$ character of the valence- and conduction-band edges. The data shown in the panels 
marked with an asterisk [$\ast$, ($T=50$~K, $x=0.23$)] have been acquired in a separate measurement on a different 
sample than used for the other temperature points at $x=0.23$. The spectra in the panels marked with a dagger [$\dag$, 
($T < 100$~K, $x=0.27$)] indicate the formation of a gap in the surface state at low temperatures as previously observed 
in Ref.~\onlinecite{Okada-Science-2013}. Further details on the latter shall be discussed elsewhere.

\begin{figure*}
\centering
\includegraphics[width=.4\textwidth]{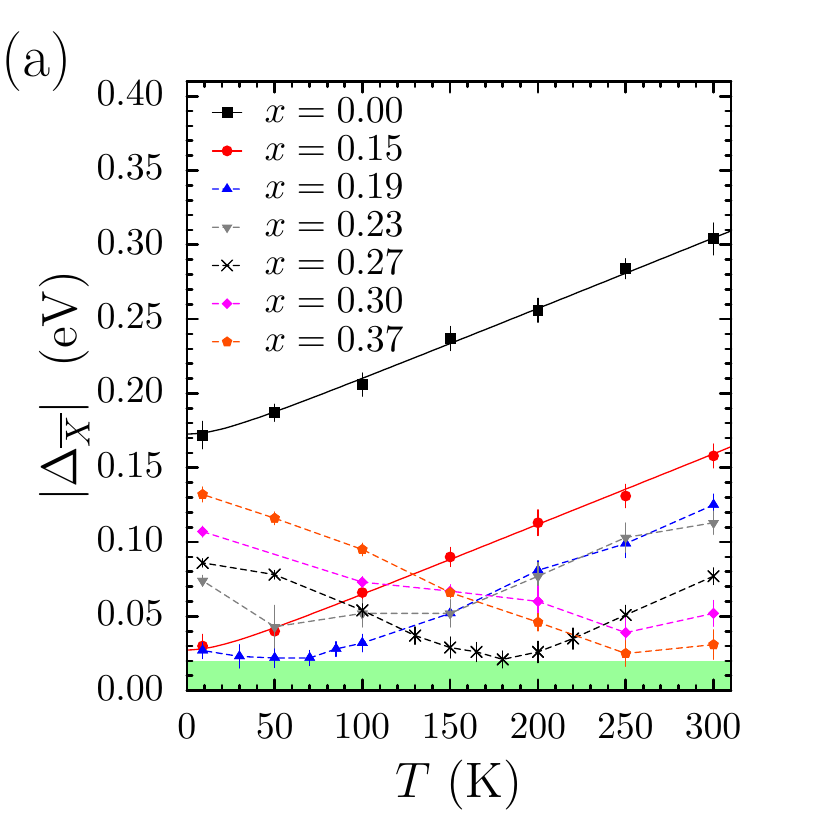}\hspace*{10mm}
\includegraphics[width=.4\textwidth]{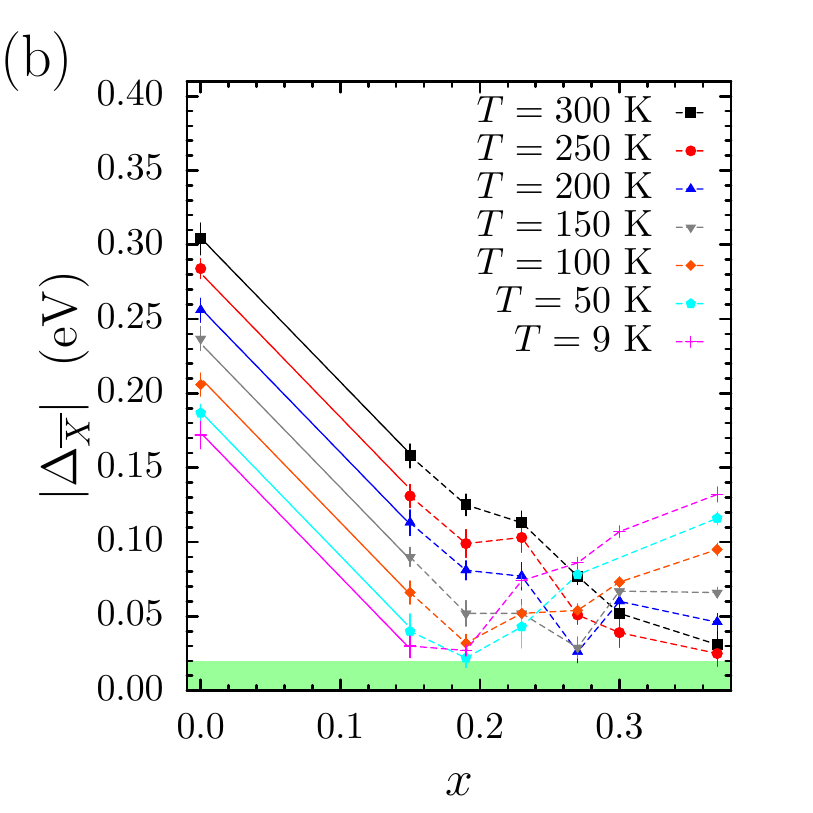}
\caption{(a) Absolute value of the energy gap at $\overline{X}$ as a function of temperature for \PbSnXSe crystals with 
different tin contents $x$. The gap values have been determined from the ARPES spectra shown in 
Figs.~\ref{fig:EDCmatrix}, \ref{fig:COLORmatrix}, and~\ref{fig:X19X27}. The green-shaded area represents the minimal 
gap that is always observed. The solid lines represent fits of Eq.~(1) to the data. The dashed lines are merely guides 
to the eye. (b) Same data as in (a) but presented as a function of $x$.}
\label{fig:absgap}
\end{figure*}

Eventually, Fig.~\ref{fig:absgap} depicts the absolute value of the determined energy gap of \PbSnXSe at 
$\overline{X}$ as a function of temperature and tin content. The data are identical to those shown in Fig.~\ref{fig:gap} 
of the main text, however, without following the convention of choosing a negative value for the gap in the 
band-inverted TCI state.

\linespread{1.0}

\begin{thebibliography}{49}
\bibitem [{Ban()}]{BandCurvature}%
  \BibitemOpen
  \href@noop {} {}\bibinfo {note} {Although the ($T=300$~K, $x=0.37$) data set
  contains an ambiguous EDC without an apparent gap close to $\overline{X}$,
  the overall band curvature indicates that the surface electronic structure is
  indeed gapped also in this case.}\BibitemShut {Stop}%
\end{thebibliography}
\end{document}